\documentclass[12pt,a4paper,fleqn]{article}

\usepackage[textheight=23cm,textwidth=15cm]{geometry}
\usepackage{amsmath}
\usepackage{graphicx}
\usepackage[american]{babel}
\usepackage{here}
\usepackage[articletitle=true]{achemso}

\usepackage{lscape}
\usepackage{longtable}

\begin{document}

\begin{center}

{\LARGE\bf
 The Transferability Limits of Static Benchmarks
}

\vspace{1cm}

{\large
Thomas Weymuth\footnote{ORCID: 0000-0001-7102-7022} and
Markus Reiher\footnote{Corresponding author; e-mail: markus.reiher@phys.chem.ethz.ch; ORCID: 0000-0002-9508-1565}
}\\[4ex]

Laboratory of Physical Chemistry, ETH Zurich, \\
Vladimir-Prelog-Weg 2, 8093 Zurich, Switzerland

May 21, 2022

\vspace{.43cm}

\textbf{Abstract}

\begin{minipage}{13cm}
{\small
Every practical method to solve the Schr\"odinger equation for interacting many-particle systems introduces approximations. Such methods are therefore plagued by systematic errors. For computational chemistry, it is decisive to quantify the specific error for some system under consideration. Traditionally, the primary way for such an error assessment have been benchmarking data, usually taken from the literature. However, their transferability to a specific molecular system, and hence, the reliability of the traditional approach always remains uncertain to some degree. In this communication, we elaborate on the shortcomings of this traditional way of static benchmarking by exploiting statistical analyses at the example of one of the largest quantum chemical benchmark sets available. We demonstrate the uncertainty of error estimates in the light of the choice of reference data selected for a benchmark study. To alleviate the issues with static benchmarks, we advocate to rely instead on a rolling and system-focused approach for rigorously quantifying the uncertainty of a quantum chemical result.
}
\end{minipage}
\end{center}

All practical quantum chemical solution methods for the many-particle (electronic and nuclear) Schr\"odinger or Dirac equation introduce certain approximations. An example for such an approximation is the introduction of a basis set (\textit{e.g.}, for the linear combination of atomic orbitals or for some superposition of Slater determinants), which has to be limited to a certain size and which is therefore finite and not complete. Since approximations introduce some error in the final result, reliable error bars for a given quantum chemical method are important for the interpretation of calculated results with confidence. However, it is usually not straightforward to quantify a method's uncertainty for a specific case under consideration\cite{Simm2017, Reiher2021, Lejaeghere2020, Rommel2021}. 

Owing to the lack of analytical results for error estimation, the reliability of quantum chemical methods is assessed by numerical benchmarking. The error with respect to some reference data is determined for a predefined set of molecules. We call this approach of preselecting a fixed set of molecules, for which reference data are provided, \textit{static} benchmarking. Savin and Pernot have recently highlighted some shortcomings of benchmark studies\cite{Savin2020}.

Naturally, if the predefined set of molecules is small, it will only be representative for a small region of chemical space. However, for any size of the set, it will be important to know (1) whether even the region of reliable applicability is contiguous at all and (2) whether the boundaries of the region can be known for some predefined accuracy required for a meaningful result. Unfortunately, such knowledge will, in general, not be accessible. Accordingly, many different options for scrutinizing approximate quantum chemical models emerged. 

Numerous numerical experiments have shown that the error of a quantum chemical result may strongly vary between different classes of compounds and even within a given class. Since a benchmark study on a small data set is not likely to be representative for the accuracy of a method across the entire chemical space, increasingly larger benchmark data sets have been proposed (\textit{e.g.}, those compiled by Curtiss \textit{et al.}\cite{Curtiss2005}, by Grimme and coworkers\cite{Goerigk2017}, by Truhlar and coworkers\cite{Peverati2014}, and by Mardirossian and Head-Gordon\cite{Mardirossian2017}), with the ultimate goal to construct reference data sets that represent molecular structures and their properties well across the entire chemical space.

The latest generation of benchmark sets, which have matured through decades of work, may be considered truly large, implying that a sufficiently large portion of chemical compound space is covered. Hence, we may subject them to statistical analysis in order to understand how conclusions regarding accuracy and transferability depend on the composition of these large benchmark sets. For example, one may expect eliminating only a single data point from any of these large sets to have a negligible effect on any conclusion. Accordingly, the utility of the set for assessing the error of a quantum chemical model theory should hardly be affected by deleting only one data point.

However, as we show in this work, even very large benchmark sets can suffer from shortcomings which prevent them from being sufficiently general for establishing reliable and transferable error bounds for quantum chemical methods. For our analysis, we chose the data set by Mardirossian and Head-Gordon\cite{Mardirossian2017} simply because of its size and the fact that it is a very recent one. We emphasize that our conclusions are likely to hold for any other choice as well as should become evident in the discussion below. Hence, they may be considered general, \textit{i.e.}, independent of the specific choice of benchmark set, as nothing in this choice for a large benchmark set creates a specific condition that would not hold for any other contemporary large basis set.

The data set compiled by Mardirossian and Head-Gordon\cite{Mardirossian2017} contains a total of 4986 data points. It comprises 84 subsets assembled from preceding benchmarking studies. It includes a broad range of energies such as absolute atomization energies, potential energy curves of rare-gas dimers, isomerization energies, non-covalent interaction energies, barrier heights, ionization potentials, and electron affinities (see also Table~1 in the supporting information). Also the molecule classes covered are rather broad, including, for example, water clusters, amino acids, graphene, alkanes, non-covalent complexes, radicals, and charged systems. However, the data set is biased towards organic molecules; the chemical elements present are limited to aluminum, argon, arsenic, boron, beryllium, bromine, carbon, chlorine, fluorine, hydrogen, helium, krypton, lithium, magnesium, nitrogen, sodium, neon, oxygen, phosphorus, sulfur, selenium, and silicon. Almost 53\,\% of all atoms are hydrogen atoms, while about 30\,\% of all atoms are carbon atoms. Naturally, one might expect a lack of transferability of benchmark results for molecules made of elements not present in this set, such as transition metal complexes and clusters. However, this aspect we do not probe because by design our analysis solely refers to elements that are contained in the data set.

\begin{figure}[H]
\begin{center}
\includegraphics[width=10cm]{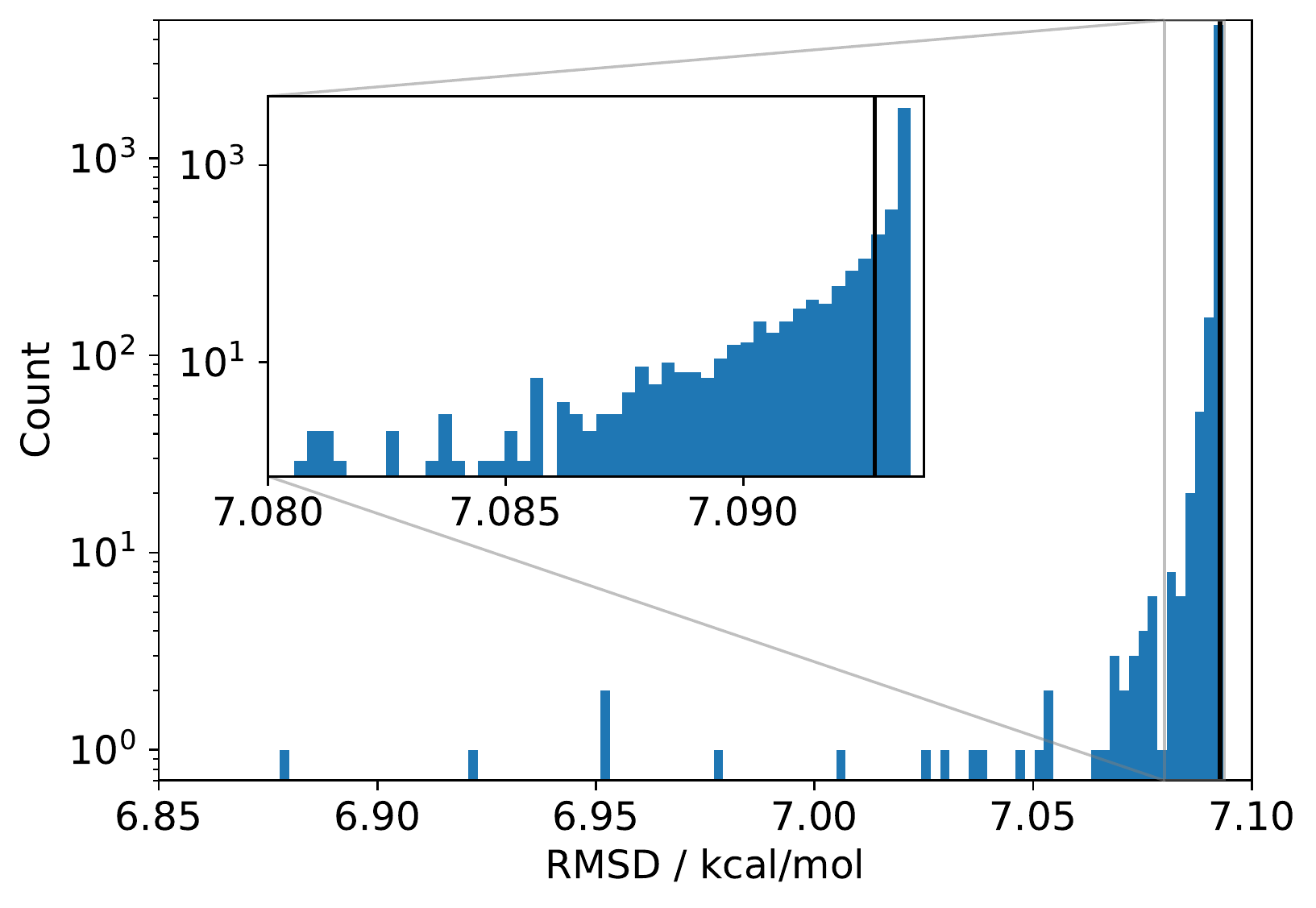}
\end{center}
\caption{\label{fig:jackknifing}\small Histogram of the RMSDs obtained for the 4986 data sets synthesized by jackknifing and evaluated with PBE data taken from Ref.~\citenum{Mardirossian2017}. The black line denotes the RMSD for PBE obtained with the original reference data set.}
\end{figure}

The data set is so vast that a manual inspection is almost impossible, and so, an in-depth analysis becomes daunting. As a starting point, we chose to carry out a simple jackknifing analysis\cite{Miller1974}, \textit{i.e.}, we created 4986 new data sets, each having exactly one point from the original data set removed. Then, we evaluated the overall root mean square deviation (RMSD) for various density functionals on these new sets (we follow Mardirossian and Head-Gordon\cite{Mardirossian2017} and take the RMSD as the primary error measure). Since the original set is larger by only one data point, we can assess how important a single data point in even a very large data set can be. The results of this jackknifing analysis for the PBE density functional\cite{Perdew1996, Perdew1997} are shown in Fig.~\ref{fig:jackknifing}. Note that the raw data was not recalculated, but was taken from Ref.~\citenum{Mardirossian2017}.

Almost all jackknife sets produce RMSDs very similar to the reference RMSD of about 7.1\,kcal/mol obtained with the original data set, which implies that the eliminated data points are well in line with this average accuracy. Note also that an RMSD of 7.1\,kcal/mol is rather large in view of chemical accuracy of about 1\,kcal/mol and points to clear deficiencies of the PBE model.

Most importantly, there are a few jackknife sets which lead to distinctively lower RMSDs, indicating that a large data set that accidentally leaves out one of these data points would assign to the PBE model a (artificially) higher overall accuracy. In the most extreme case, removing just a single data point lowers the overall RMSD to about 6.9\,kcal/mol, which is about 3\,\% smaller than the reference RMSD. Although an average error lowered by 0.2\,kcal/mol seems to be not particularly large in absolute terms, it is nevertheless astonishing that removing a single data point from a very large set can have such a large effect.

Moreover, removing the data point with the largest error does not necessarily have the same relative effect for all functionals. That is, for a better functional with a smaller overall RMSD, removing this point can have a much larger effect than for a less accurate functional such as PBE. For example, B97M-rV\cite{Mardirossian2017b}---according to the study by Mardirossian and Head-Gordon the leading functional in the class of functionals lacking exact exchange---features an overall RMSD of 3.1\,kcal/mol, \textit{i.e.}, its RMSD is only half as large as that of PBE. When removing the point with the largest error, the RMSD drops also by about 0.2\,kcal/mol, the same amount as in the case of PBE. Hence, in this case, the RMSD can be ``improved'' by about 6\,\% by simply leaving out a single data point.
Note that for both functionals, the data point with the largest error is the same one, namely data point no.~18 out of the AE18 subset; for PBE, the absolute error is 122\,kcal/mol, while it is 79\,kcal/mol for B97M-rV (see also below).

It is not surprising that the lowest RMSD is obtained for the jackknifed set which does not contain the data point with the largest error. However, this may also be taken as a recipe for producing a large data set that will assign an artificially high accuracy to the model. Building upon this idea, we can construct a new data set by removing the ten data points with the largest errors. Even though this data set is almost as large as the original one, it leads to an RMSD which is now 17\,\% lower than the reference RMSD for PBE.
Even more striking is the case of
B97M-rV, for which leaving out the ten data points with largest errors (for B97M-rV)
lowers the RMSD by 31\,\%. These results highlight the pronounced effect that a few individual data points can have on the overall measure for accuracy and reliability, in our case on the RMSD. 

However, one might argue that it is in fact our analysis which is artificial as one would not be allowed to remove data points of high error in a rigorous study. Still, our argument is that this could happen accidentally and might even be the case for the original data set, for which it is not clear whether reference data with a large error might not even be present because such reference data was simply not available. A trivial example is an application to a chemical system which is not even well represented by this reference set such as a transition metal system (\textit{cf.}~the examples in the WCCR10 set of ligand dissociation energies\cite{Weymuth2014, Husch2018}).

To conclude, if, by accident, one would have constructed the large data set without these ten data points, which is a tiny amount of data compared to the total number of data points, one would have come to the conclusion that the accuracy of PBE is higher than currently believed. Or, to turn this argument around, new data points can reduce or increase the currently assessed accuracy of a density functional up to 20\,\%.

It is no surprise then that also the relative ranking of density functionals can change upon leaving away only a few data points. With the original data set, PBE is ranked 164th, while MN15\cite{Yu2016} is ranked 14. Upon removal of the ten points with largest error for PBE from this data set, PBE improves to rank 161. Also MN15 improves, raising to rank 9, \textit{i.e.}, it improved even more than PBE. In some cases, the change in relative ranking can be very pronounced. For example, B97M-rV\cite{Mardirossian2017b} improves from rank 31 to 10, the LC-VV10 functional\cite{Vydrov2010} even rises from rank 150 to 93, and SOGGA11-X\cite{Peverati2011} falls from rank 44 to 64.

\begin{figure}[H]
\begin{center}
\includegraphics[width=10cm]{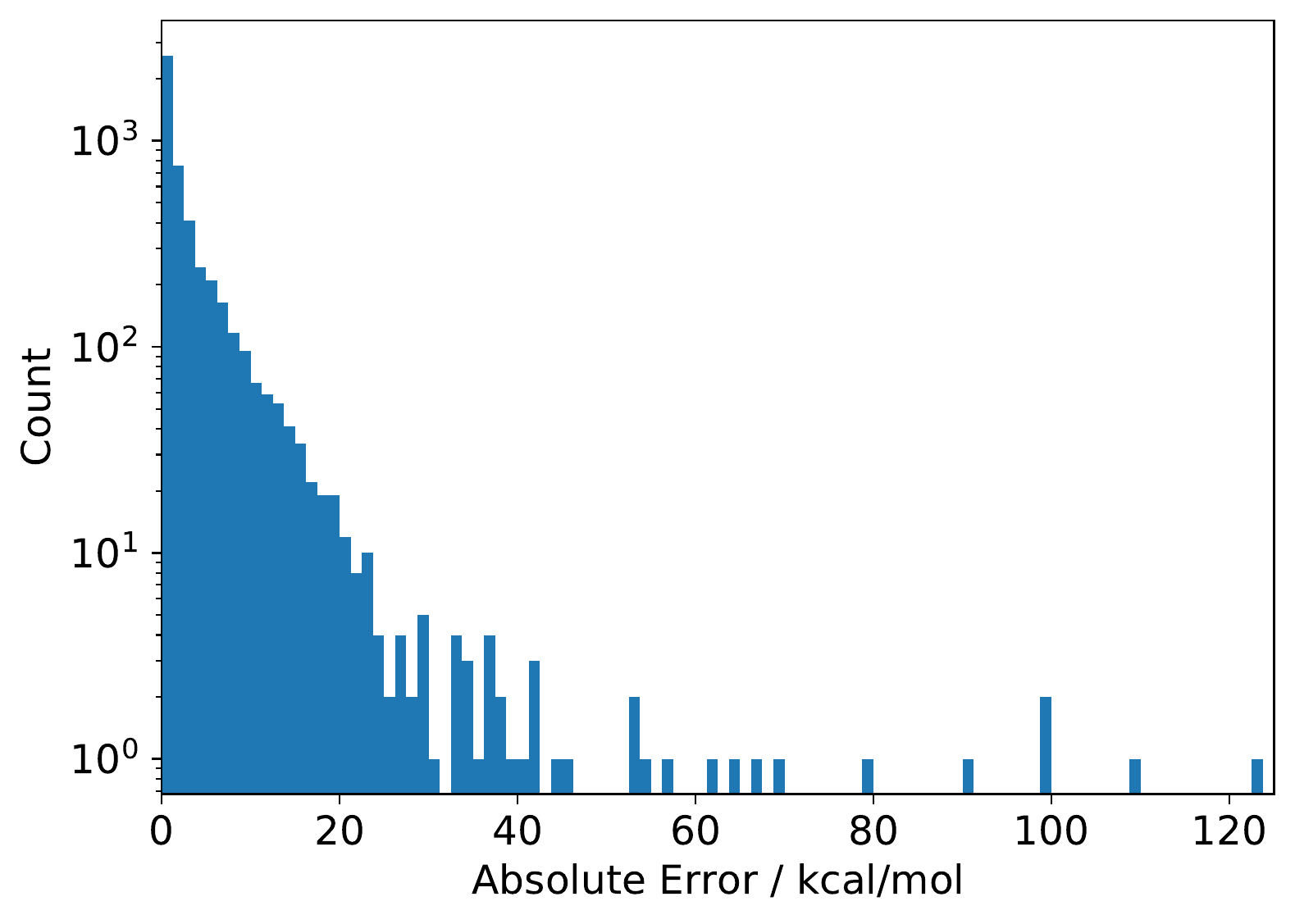}
\end{center}
\caption{\label{fig:errors}\small Histogram of the absolute errors of the PBE functional measured as the absolute differences between the reference values and the corresponding values calculated with PBE. The data depicted here were obtained from the original reference data set
in Ref.~\citenum{Mardirossian2017}}
\end{figure}

It is instructive to investigate the distribution of the absolute error of the individual data points contained in the original data set shown in Fig.~\ref{fig:errors}.
The absolute error is calculated as the absolute difference between the reference value and the corresponding value calculated with a given density functional such as PBE. We see that almost all absolute errors are below 20\,kcal/mol. However, a few errors are much larger, with the largest one reaching a huge value of more than 120\,kcal/mol (see below for more details). Clearly, this data point strongly reduces the RMSD when left out, which is exactly what we observed as its omission led to the smallest RMSD in Fig.~\ref{fig:jackknifing}. In fact, we see a complementarity of the distribution of RMSDs in Fig.~\ref{fig:jackknifing} and the error distribution in Fig.~\ref{fig:errors}---they are almost mirror images of one another.

Inspection of Fig.~\ref{fig:errors} prompts one to consider the large-error data points as outliers, unnecessarily skewing the error distribution. However, the fact that there are only so few data points with errors larger than 60\,kcal/mol might also simply be due to a general scarcity of reliable reference data. In this case, however, one could argue that this error range is underrepresented in the data set. Since adding or leaving out a few points in this range has a large effect on the overall error measures, the current error measures are likely not to be indicative of the ``true'' error measures which one would obtain when properly representing the entire error range. Hence, even when using a very large data set for static benchmarking, it is possible that the error measures obtained on this data set are not representative of the true error a given functional (or, in fact, any approximate quantum chemical method) exhibits.

Furthermore, considering the broad range spanned by the absolute errors in Fig.~\ref{fig:errors}, it is also clear that a single number (here, the RMSD) does not carry enough information to truly reflect the accuracy and reliability of a given density functional for a specific purpose. At the very least, the minimal and maximal errors (or some measure for the distribution of errors such as the standard deviation) are to be considered as well, which is why typically the spread (\textit{e.g.}, measured in terms of a standard deviation) or the largest absolute error are reported as well. An alternative method to deal with benchmark sets leading to errors of largely different size is to introduce (arbitrary) scaling factors reducing large errors and increasing small errors, leading to a narrower error distribution. This is the case, for example, for the GMTKN55 database by Grimme and coworkers\cite{Goerigk2017} (see also below for a more detailed discussion of this approach.)

Naturally, absolute error measures such as the RMSD can be problematic when a data set combines points of largely different magnitudes. As we have seen, the largest absolute error in the data set of Mardirossian and Head-Gordon amounts to about 120\,kcal/mol (for PBE). This data point belongs to the AE18 subset which contains absolute atomization energies, \textit{i.e.}, a quantity which can easily reach large absolute values. The value for this data point obtained with PBE is $-$330'913.7503\,kcal/mol, which is three orders of magnitude larger than typical reaction energies. Therefore, an error of 120\,kcal/mol represents a relative error of only about 0.04\,\%. In contrast to this, there are data points with a much smaller absolute error but a much larger relative error. Therefore, relying only on absolute error measures can be misleading for judging the accuracy of a given density functional. Specifying not only an absolute measure, such as the RMSD, but also a relative error measure yields more insight and, hence, a better informed decision when choosing an exchange--correlation functional.

\begin{figure}[H]
\begin{center}
\includegraphics[width=15cm]{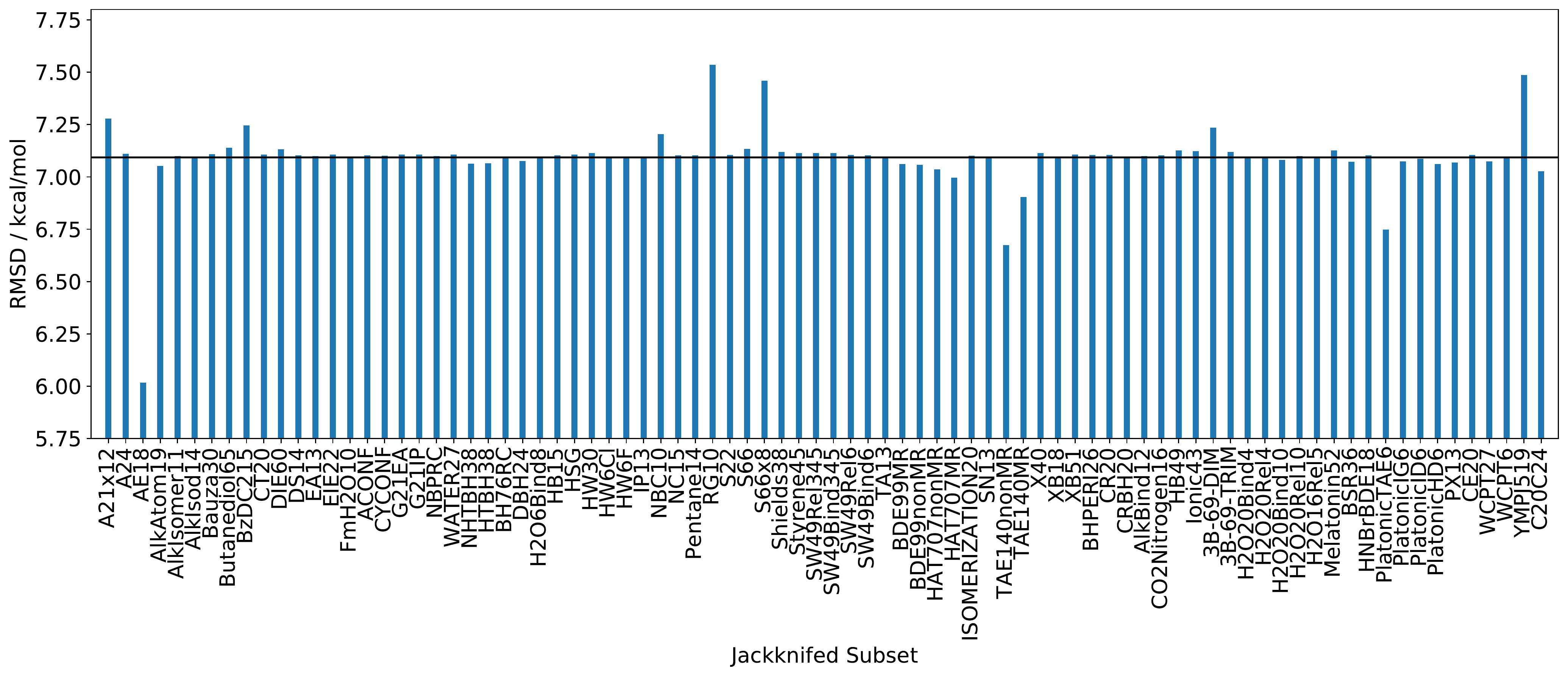}
\end{center}
\caption{\label{fig:sets_jackknifed}\small RMSDs of a subset-jackknifing analysis for the PBE density functional. Abscissa: eliminated subsets named as in Ref.~\citenum{Mardirossian2017}. The horizontal black line denotes the RMSD for PBE obtained with the original (complete) data set.}
\end{figure}

Given the fact that the whole benchmark set was built of a collection of subsets raises the question of how relevant an individual class of data points is. Hence, we may investigate the jackknifing of entire subsets next, \textit{i.e.}, instead of omitting individual points, we leave out entire subsets. The result of such an analysis is shown in Fig.~\ref{fig:sets_jackknifed}.

For most subsets, leaving any of them out hardly affects the RMSD. However, a few subsets have an unexpectedly large effect. As one would have expected following the discussion so far, eliminating the AE18 subset\cite{Chakravorty1993} of absolute atomization energies reduces the RMSD most, \textit{i.e.}, by more than 1\,kcal/mol, which is about 15\,\% of the original RMSD for PBE. Note that the AE18 subset consists of only 18 data points. By contrast, omitting the entire HAT707nonMR subset\cite{Karton2011} with its 505 data points has a negligible effect on the overall RMSD. Hence, also at the level of individual subsets, we observe that some sets of data points have a disproportionately large effect, again pointing to systematic problems that certain parametrized models will face if different classes of physico-chemical properties are combined in one benchmark set.

It is instructive to analyze the individual subsets and their effect on the overall RMSD in more detail. In the supporting information, Tables~1 and~2 list all subsets contained in the data set of Mardirossian and Head-Gordon, a short description of the type of data and how many data points they contain. These tables in the supporting information also collect ``uncertainty labels'' (UL) for selected exchange--correlation functionals. Each of these uncertainty labels is composed of two values. The first one specifies the difference between the reference RMSD (obtained on the full data set) and the largest RMSD obtained when eliminating a single data point from this subset (given in percent of the reference RMSD). The second value is the RMSD difference obtained when leaving out the entire subset (again in percent of the reference RMSD).

Irrespective of the specific density functional, we find that there are three distinct groups of ULs. First, there are subsets for which the absolute values of both components of the UL are very small, \textit{e.g.}, H2O16Rel5, HSG, and SN13 (for the PBE density functional). These are subsets in which all data points have rather small errors. Hence, eliminating any one of them from the benchmark data set hardly affects the overall RMSD. These subsets are also all rather small, having always less than 100 data points (\textit{i.e.}, less than 2\,\% of the overall data set size), which explains why the omission of the entire subset also has a negligible effect on the overall RMSD.

The second group of subsets are those for which the first number of the UL is exactly 0.0\,\%, while the second number is rather large, \textit{i.e.}, larger than 1.0\,\%. For PBE, these sets are 3B-69-DIM, A21x12, BzDC215, NBC10, RG10, S66x8, and YMPJ519. Obviously, all data points in these subsets have rather small absolute errors so that omitting any single point does not affect the overall RMSD at all. Interestingly, according to Fig.~\ref{fig:sets_jackknifed}, we find that for exactly these subsets, their omission leads to a significantly larger RMSD, as reflected in the second component of the UL. This is caused by the fact that all these subsets are rather large, comprising between 184 and 569 data points. Therefore, ignoring such an entire subset removes a significant part of the overall data set. This large part, when present, reduces the overall RMSD because all points in these subsets have very small errors (as shown by the first component of the UL). Therefore, omission of any of these subsets will lead to an increased overall RMSD.

Finally, there is a third group of subsets having ULs in which the second component is negative and rather large in absolute terms and the first number is different from zero (in some cases also being rather large in absolute terms). These subsets are AE18, HAT707MR, TAE140nonMR, TAE140MR, and PlatonicTAE6 (again, for PBE). These are exactly the sets which contain data points with a large absolute error. Leaving out such types of data points will have a notable effect on the overall RMSD, as exemplified by the first UL component being nonzero. Omitting such an entire subset will only increase this effect. Therefore, the second UL component is even larger. Note also that these subsets are not necessarily large; for example, PlatonicTAE6 contains only 6 data points. 

Compared to the second group of subsets, for this third group the large second part of the UL is \textit{not} due to the fact that many points with a small error are removed, but because a few points with a large error are removed. This is also why removing the subsets from the third group decreases the overall RMSD (\textit{cf.}, Fig.~\ref{fig:sets_jackknifed}), whereas it is increased for the second group of subsets.

For the sake of completeness, one should note here that the subset HAT707nonMR is an interesting exception, at least for PBE. With 505 data points, it is one of the largest subsets, yet Fig.~\ref{fig:sets_jackknifed} shows that removal of this subset neither significantly decreases or increases the overall RMSD. The first part of the UL is not exactly 0.0\,\%, explaining why this subset, despite its size, cannot belong to the second group of subsets mentioned above: obviously, there are a few data points in HAT707nonMR which are comparatively large, so that removal of one of them is already visible in the overall RMSD. However, these errors are not so large that a removal of the entire subset would significantly decrease the overall RMSD. This is reflected in the second part of the UL, the absolute value of which is not that large corresponding to $-$0.8\,\%. Still, it is non-negligible, and in fact, it is very large for other density functionals. Therefore, HAT707nonMR is a fringe case for the PBE functional, but for other exchange--correlation density functional approximations it would be attributed to the third group of subsets.

This prompts us to consider a comparison of a few representative density functionals. To this end, we assembled the ULs of PBE\cite{Perdew1996, Perdew1997}, B97-D3(0)\cite{Grimme2006}, TPSS\cite{Tao2003}, B97M-rV\cite{Mardirossian2017b}, PBE0\cite{Adamo1999, Ernzerhof1999}, and $\omega$B97M-V\cite{Mardirossian2016} in Tables~1 and~2 in the supporting information. These functionals span several rungs of Jacob's ladder; PBE and B97-D3(0) are GGA functionals, TPSS and B97M-rV are meta-GGA functionals, while PBE0 and $\omega$B97M-V are hybrid functionals. PBE, TPSS, and PBE0 are nonempirical functionals, \textit{i.e.}, they contain no empirical parameters. This lack of ``explicit empiricism'' makes them appealing from a fundamental point of view. Moreover, they are readily available in many quantum chemistry computer program packages. B97-D3(0), B97M-rV, and $\omega$B97M-V have been chosen since these are, according to the study by Mardirossian and Head-Gordon\cite{Mardirossian2017}, the best functionals of their respective rung on Jacob's ladder. These are all empirical functionals, having 11 (B97-D3(0)) and 12 (B97M-rV and $\omega$B97M-V) parameters. Therefore, judging from the number of parameters, these three functionals are all comparable; if one performs significantly better than the others, this must therefore be due to some intrinsic advantage of this functional, rather than simply because of an increased flexibility owing to a larger number of parameters.

It is important to stress that a comparison of density functionals according to the ULs is not at all the same as comparing them according to their RMSD. Rather than being a direct measure of the error (such as the RMSD), the ULs of a functional are a measure for the uncertainty of the error of a density functional as obtained from a certain benchmark set. A truly reliable functional is not only expected to have a low overall error, but also a small uncertainty in this very error. Hence, a reliable performance analysis of exchange--correlation functionals (or any other physico-chemical model) should take into account ULs or a similar measure.

Overall, the ULs show the same general trends for all density functionals. For almost all subsets, the ULs of each exchange--correlation functional belong to the same group (out of the three groups identified above). As already observed by Mardirossian and Head-Gordon\cite{Mardirossian2017}, functionals higher up on Jacob's ladder generally have a better overall performance. However, this is not consistent across all subsets. For example, for the C20C24 subset (containing isomerization energies of the ground state structures of C\textsubscript{20} and C\textsubscript{24}), PBE
(on the second rung of Jacob's ladder; UL $-$0.6\,\%, $-$0.9\,\%) is significantly better than B97-D3(0) (second rung; UL $-$4.1\,\%, $-$9.1\,\%), TPSS (third rung; UL $-$2.3\,\%, $-$3.7\,\%), B97M-rV (third rung; UL $-$3.3\,\%, $-$5.2\,\%), and $\omega$B97M-V (fourth rung; UL $-$0.7\,\%, $-$1.5\,\%), and only slightly worse than PBE0 (fourth rung; UL $-$0.5\,\%, $-$0.8\,\%).

Moreover, the best-on-rung functionals B97-D3(0), TPSS, and $\omega$B97M-V are not always better than the other functionals on the same rung of Jacob's ladder. The C20C24 subset is again a good example: B97-D3(0) is significantly worse than PBE, B97M-rV is clearly worse compared to TPSS, and also $\omega$B97M-V performs worse than PBE0. However, the opposite observation can also be made, most importantly for the AE18 subset. Here, B97-D3(0), B97M-rV, and $\omega$B97M-V are clearly superior to PBE, TPSS, and PBE0. In summary, we understand that it is not obvious at all that density functionals can be ranked in a general sense. A given functional may excel for a specific type of physico-chemical property, while its performance may be deteriorating for another. This observation relates to the approximate nature of the density functionals, which affects different properties differently.

When considering the third group of subsets identified above, we realize that, with the exception of HAT707MR, all of them contain atomization energies. This is an extensive quantity, \textit{i.e.}, one which increases with increasing molecular size. Also the errors of such atomization energies are dependent on the molecular size, being larger for bigger molecules. This is another example highlighting the limited transferability of static benchmark results obtained for a predefined set of molecules. Depending on the actual size of the molecules in this predefined set, the reported error might be larger or smaller, not necessarily reflecting the error resulting in a given application. Of course, a straightforward way to circumvent this particular problem of size-extensive errors is to normalize all errors to the molecule size.

However, other transferability issues are not so easy to address. Consider, for example, isodesmic reaction energies, an intensive quantity. Such reactions are deliberately set up to exploit error compensation as much as possible. Hence, a hypothetical benchmark conducted on a set of such isodesmic reaction energies is likely to yield an error measure which is lower than one observed for some other application. While it is clear from the outset that an informed error measure for a certain method cannot be achieved by considering isodesmic reactions alone, it is not at all clear what set of properties (and which molecules) has to be considered in order to achieve such a well-informed error measure.

As has been mentioned above, some benchmark sets such as GMTKN55\cite{Goerigk2017} introduce arbitrary scaling factors to narrow the error distribution, that is, small errors are scaled up (\textit{e.g.}, by a factor of 10 according to the weighting scheme ``WTMAD-1'' of GMTKN55\cite{Goerigk2017}) while large errors are scaled down (for example, by a factor of 10 in ``WTMAD-1''). Disregarding the fact that there is a rather large degree of arbitrariness in the specific choice of such scaling factors, we would like to emphasize that such an approach is fundamentally flawed. This is because it artificially enhances the influence of certain data points, while the role of others is decreased. Moreover, if one is interested in, say, absolute atomization energies for a specific application, one will be interested in the true error of a certain model for evaluating atomization energies and not in some scaled value of the true error.

Even if it were possible to adequately represent the entire chemical space by a single benchmarking set, aggregating all the errors into one overall error measure would lead to this error measure being too high for some parts of chemical space (where the applied model is particularly accurate) and too low for other parts. This holds also true if individual properties are studied separately (as done, \textit{e.g.}, by Mardirossian and Head-Gordon\cite{Mardirossian2017}), because even for the same physico-chemical property, the errors of a quantum chemical method are distributed heteroscedastically, \textit{i.e.}, not evenly, across chemical space. Therefore, a huge challenge for static benchmarking is the fact that the intrinsic errors of any quantum chemical method are usually not distributed evenly across chemical space\cite{Simm2017}.

While compiling our results here, we noted that Gould and Dale very recently arrived at the same conclusion\cite{Gould2022}. To solve the problem that a single overall error measure obtained from a large benchmark set can mask systematic deficiencies of a given density functional, Gould and Dale proposed new benchmark sets comprising so-called ``poison'' reactions, \textit{i.e.}, reactions which are known from experience to be difficult to model accurately with many exchange--correlation functionals. An error measure obtained from these new benchmark sets will indeed be more representative for the specific reactions contained in these sets. However, it will not be any more transferable than statistical measures obtained from any other benchmark set; for many applications, a given density functional might perform much better than suggested by the errors obtained from a particularly difficult to model benchmark set. The heteroscedasticity of the errors of quantum chemical methods is a fundamental challenge to static benchmarking. It implies that results of static benchmarking are of limited transferability; this problem cannot be solved by any particular principle for the construction of static benchmark sets.

A simple yet effective way to overcome all of these challenges of static benchmarking is to adopt rolling system-focused benchmarking\cite{Simm2016, Simm2018, Proppe2019a, Reiher2021}. By comparing against reference data for exactly these molecules in which one is interested, one can make sure to obtain reliable performance metrics, albeit only for the part of chemical space covered. As the focus is expanded to other molecules, these are incorporated dynamically into the benchmark set to make sure that the error assessment is accurate also for these new molecules. And naturally, one may even remove information from the increasing benchmark data set in a system-focused parametrization should the parametrized model turn out to be too unflexible in its analytical form to accommodate the full freedom of an exact first-principles description.

It is important to stress that, while a system-focused benchmarking approach guarantees to yield error measures relevant for the specific application under consideration, these error measures will still be affected by some uncertainty themselves. This uncertainty could be estimated with an approach such as bootstrapping\cite{Efron1979}. For a very accurate (but still not fully exact) model, it is possible that the uncertainty of the error measure is of the same order of magnitude as the error measure itself---and clearly, more work on Bayesian error estimators in the context of quantum chemical approaches will be needed to identify the most reliable ones.

A dynamic benchmarking approach lends itself naturally to applications which continuously produce data such as high-throughput virtual screening and \textit{ab initio} molecular dynamics. It is especially favorable for the exploration of vast chemical reaction networks\cite{Unsleber2020}. On the one hand, such networks are naturally suited to a rolling benchmarking, as the exploration itself proceeds in a rolling fashion. On the other hand, one can easily imagine such networks to cover parts of chemical space for which no reliable reference data is available yet. 

Due to the heteroscedastic nature of the error of quantum chemical methods, it is impossible in such a situation to provide reliable error bars relying only on existing reference data. Of course, the calculation of accurate new reference data is time-consuming and may pose prohibitive barriers in terms of computational feasibility. However, this bottleneck can be circumvented by adopting counter measures. For instance, in a two-step approach to uncertainty quantification, starting from a rather small set of compounds for which it is possible to calculate accurate reference data, a machine-learning model can be trained to predict the error. This model is then applied to all new molecules added to the set or network. Crucially, the machine-learning model needs to provide an uncertainty measure for the predicted errors, such that it becomes obvious when the machine-learning predictions become too inaccurate. We have demonstrated such an approach in Refs.~\citenum{Simm2018} and~\citenum{Proppe2019a}.

\section*{Acknowledgments}
\label{sec:acknowledgments}

This work was generously supported by the Swiss National Science Foundation (SNSF) through project no.~200021\_182400.

\providecommand{\latin}[1]{#1}
\makeatletter
\providecommand{\doi}
  {\begingroup\let\do\@makeother\dospecials
  \catcode`\{=1 \catcode`\}=2 \doi@aux}
\providecommand{\doi@aux}[1]{\endgroup\texttt{#1}}
\makeatother
\providecommand*\mcitethebibliography{\thebibliography}
\csname @ifundefined\endcsname{endmcitethebibliography}
  {\let\endmcitethebibliography\endthebibliography}{}

%
%
%

\end{document}